\documentclass[amsmath,amssymb,twocolumn]{iopart}
\usepackage{graphicx}
\usepackage{dcolumn}
\usepackage{bm}

\begin{document}

\title{Can nanotechnology experimentally solve the plane-plane challenge?}
\author{Alessandro Siria$^{1,2}$, Serge Huant$^1$, Geoffroy Auvert$^{3}$, Fabio Comin$^4$ and Joel Chevrier$^{1}$}
\address{$^{1}$ Institut N\'eel, CNRS and Universit\'e Joseph Fourier Grenoble, BP 166 38042 Grenoble Cedex 9, France\\
$^{2}$ CEA/LETI-MINATEC, 17 Avenue des Martyrs 38054 Grenoble Cedex 9, France\\
$^{3}$ ST Microelectronics, 850 rue Jean Monnet 38926 Crolles, France\\
$^{4}$ ESRF, 6 rue Jules Horowitz 38043 Grenoble Cedex 9, France}

\begin{abstract}
Non-contact interaction between two parallel flat surfaces is a central paradigm in sciences. This situation is the starting point for a wealth of different models: the capacitor description in electrostatics, hydrodynamic flow, thermal exchange, the Casimir force,  direct contact study, third body confinement such as liquids or films of soft condensed matter. The control of parallelism is so demanding that no versatile single force machine in this geometry has been proposed so far. Using a combination of nanopositioning based on inertial motors and microcrystal shaping with Focused Ion Beams (FIB) we propose here an experimental set up that should enable one to measure interactions between movable surfaces separated by gaps in the micrometer and the nanometer ranges.
\end{abstract}

\maketitle

\begin{figure}[t]
\center
\includegraphics[width=0.85\columnwidth]{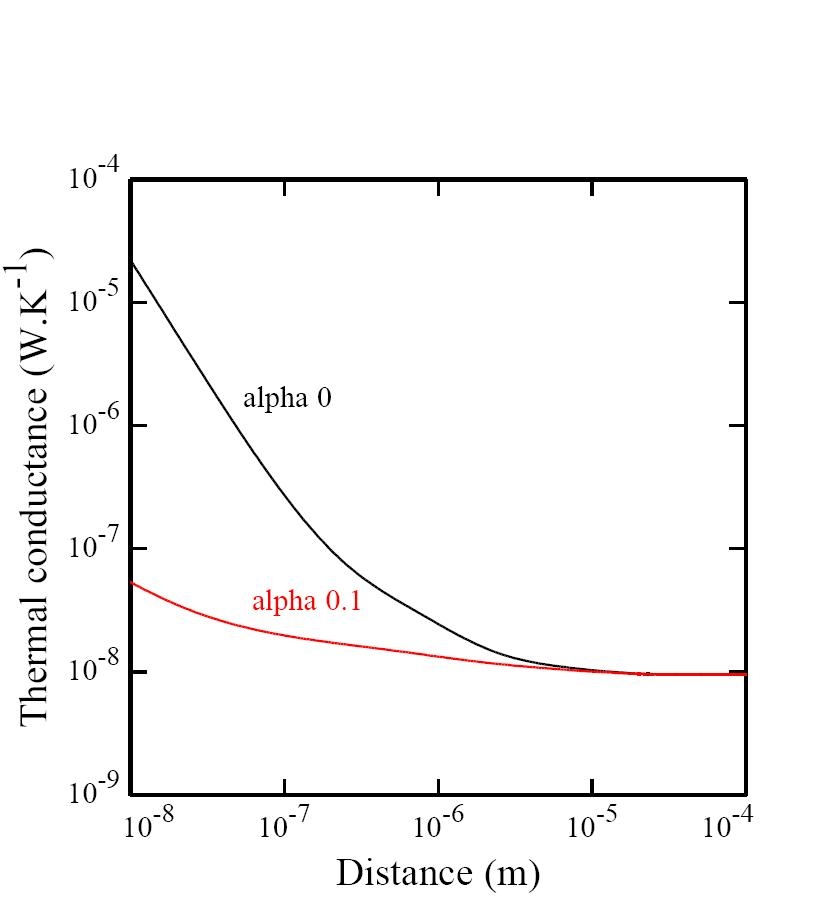}\\
\caption{\label{fig:simul} Radiative heat transfer between two flat silicon samples. The black curve (alpha 0) is for two perfectly parallel planes; the red curve (alpha 0.1) is for two planes with a residual misalignment of $10^{-1}$ $\deg$.}
\end{figure}

Measurement of non-contact interactions between surfaces has always been a challenge. This includes the presence of a third body (e.g. liquid or gas) in the separating gap, with gap varying from micrometres down to the nanoscale. In the context of complex fluids, precise measurements between extended and curved surfaces have been done using the Surface Force Apparatus (SFA) \cite{Charlaix1, Lamo, Caru}.\\
In any environment, such as vacuum, gas and liquid, the measurement of interactions between nanoobjects has greatly benefited from the Atomic Force Microscope (AFM) \cite{Mohi, Guillaume, Cap1, Cap2, Cap3, NatPhot}. Control and resolution in AFM interaction measurement has reached picoNewton scale in surface imaging under ultra-high vacuum. Cryogenic environment for single electron spin detection pushed the limit down to the attoNewton scale \cite{burg07}.\\
Measurements in the plane-plane geometry have been attempted in the framework of Casimir force studies, in order to probe the theoretical predictions of mechanical effects of quantum vacuum fluctuation \cite{Lamo, Caru, Casimir}. The limited precision obtained on historical Casimir force measurements based on macroscopic setup controls shows the extraordinary difficulty of this measurement (the reported agreement between theory and experimental data in \cite{Caru} is $\approx15\%$).\\
The use of the plane-plane geometry remains somewhat as a dream as this is the simplest geometry used in many models. Indeed, apart from controllable side effects, it enables one to make exact calculations.\\
The commonly used geometry for experiments is the sphere-plane geometry \cite{Mohi, Guillaume, Cap1, Cap2, Cap3, NatPhot}. This immediately implies the use of Derjaguin approximation often referred as Proximity Force Approximation (PFA) whose validity has always been  a matter of endless debates \cite{PFA}.\\
Beside the fact that measurements made in plane-plane geometry best compared with ideal configuration explored by theory, flat surfaces are also easier to control at nanoscale (roughness, contamination, chemical functionalization, surface patterning).\\
Previous attempts for plane-plane interaction force instruments were closely related to the SFA \cite{Lamo, Caru} .  The misalignment between the planes surfaces were controlled using capacitive force between two rotating metallic plates related to the samples. The precision allowed by this kind of set-up on the angle control was $\approx\ 2\times10^{-3}\ \deg$ \cite{Caru}.\\
An unprecedented precision in the parallelism between plane surfaces has been achieved by the Nesvizhevsky's group at the Institut Laue Langevin \cite{valery} . During the measurement of the quantum states of neutrons in the Earth gravitational field, these authors were able to align two macroscopic planes (10 cm size) with an angular precision of $\approx\ 10^{-4}$ $\deg$. However none force measurement set-up has been implemented in their experiment.\\
In this paper we analyse how the combined use of (i) inertial motors for nanopositioning (translation and rotation), (ii) nanotools such as FIB and (iii) X-ray diffraction on single crystal for real time, \textit{in situ} alignment control can overcome the key difficulties in the design of a plane-plane Surface Force Machine (p2SFM). We shall see in (i) that inertial motors originally designed for low-temperature scanning-probe microscopy (SPM) (see \textit{e.g.} \cite{canaille, biscuit}) are used both to control and vary the distance \textit{d} between interacting surfaces and their relative orientation ($\theta$, $\phi$), in (ii) FIB is used to precisely weld a flat silicon single crystal at the extremity of an AFM cantilever and  in (iii) the sharpness of X-ray diffraction at Bragg position is sufficient to control parallelism and that it can be implemented so that this control takes place \textit{in situ} and in real time.\\ 
When surfaces are kept at micron or sub-micron distances from each other, interaction phenomena, generally neglected at the macroscale, take placed between them. At the submicron scale major interaction forces between surfaces are:
\begin{itemize}
\item Electrostatic forces;
\item Hydrodynamic forces mediated by the confined fluid environment;
\item Near-field radiative heat exchanges;
\item Van der Waals and Casimir forces.
\end{itemize}
If we consider the plane-plane configuration detailed above, it is possible to define the dependence of the various interactions on the distance $d$ between surfaces. Choosing to list the interactions from the weakest to the strongest dependence on distance, we have:
\begin{itemize}
\item Hydrodynamic force (perfect slip boundary conditions) \cite{PRL, tabeling, drezet}: 
\begin{equation}
F=-\gamma\cdot v=-\frac{2\eta Av}{d}\Rightarrow\ F\rightarrow 1/d,
\end{equation}
with $A$ the interacting surface, $\eta$ the fluid viscosity and $v$ the plate velocity;
\item Electrostatic force between two conductors: 
\begin{equation}
F=-\frac{1}{2}\frac{\varepsilon V^2A}{d^2}\Rightarrow\ F\rightarrow 1/d^2,
\end{equation}
with $V$ the voltage drop between the conductors, $A$ their interacting surface, and $\varepsilon=\varepsilon_r\varepsilon_0$ the medium permittivity;
\item Radiative heat transfer between dielectric materials \cite{rousseau, mullet, ref13, ref14, ref15, ref16, ref18}: 
\begin{equation}
\phi\rightarrow 1/d^2,
\end{equation}
\item Hydrodynamic force (no slip boundary conditions) \cite{tabeling, drezet}: 
\begin{equation}
F=-\gamma\cdot v=-\frac{\eta wL^3}{d^3}\Rightarrow\ F\rightarrow 1/d^3,
\end{equation}
with $w$ and $L$ the dimensions of the plate, $\eta$ the fluid viscosity, and $v$ the plate velocity;
\item Casimir force between two perfect mirrors ($\varepsilon=-\infty$) \cite{Casimir}:
\begin{equation}
F=-\frac{\hbar c\pi^2A}{240d^4}\Rightarrow\ F\rightarrow 1/d^4.
\end{equation}
with $\hbar=h/2\pi$ the Plank's constant, $c$ the speed of light in vacuum and $A$ the plate surface.
\end{itemize}
In a recent paper \cite{PRL} we have presented a comparison between experiments and theory for the hydrodynamic force with perfect slip boundary conditions \cite{drezet} between a flat AFM cantilever and a plane substrate. The cantilever oscillated in air while the plane substrate was approached from hundreds microns down to hundreds nanometers. The plane substrate was mounted over a three linear inertial motors based on stick-and-slip technology allowing a millimeter range displacement (8 mm) with nanometer scale resolution ($\approx$ 40 nm/step). The agreement between the experimental data and the theoretical model obtained solving Navier-Stokes equation together with perfect slip boundary conditions was 5 $\%$, when a residual misalignment of 0.7 $\deg$ is considered. The experimental set-up developed for the measurement did not present any misalignment correction. However the residual misalignment between the two planes does not preclude the comparison between theory and experiment. This is because of the weak distance dependence of the interaction force in analysis.\\
A simulation of the influence of the misalignment can be done in the case of the radiative heat transfer. We can consider, for example, the case of two flat surfaces of p-doped silicon ($n\approx 5\times10^{18}\ cm^{-3}$) \cite{rousseau, mullet, ref13} and compute the thermal conductance between them as done in Fig. \ref{fig:simul}. We can see that a control in the parallelism much better than  $10^{-1}$ $\deg$ is needed for a reliable comparison between theory and experiments at submicron scales. In the case of Casimir force the required angle control is still more demanding. From Bressi \textit{et al.} \cite{Caru} one sees that a precision better than $10^{-3}$ $\deg$ is needed for a comparison between experiments and theory much better than 15 $\%$. Such a level of accuracy cannot be achieved using a static experimental set-up as in the case of the hydrodynamic force measurement. An experimental set-up allowing for a real time \textit{in situ} correction of the misalignment has to be implemented in the force machine.

\begin{figure}[t]
\center
\includegraphics[width=0.55\columnwidth]{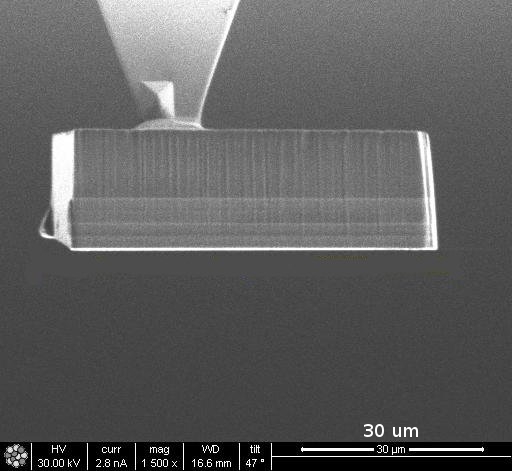}\\
\caption{\label{fig:fib} Scanning electon micrograph of a sample realised using FIB milling. A cubic like block has been welded at the end of an AFM cantilever.}
\end{figure}

\begin{figure}[ht]
\center
\includegraphics[width=0.85\columnwidth]{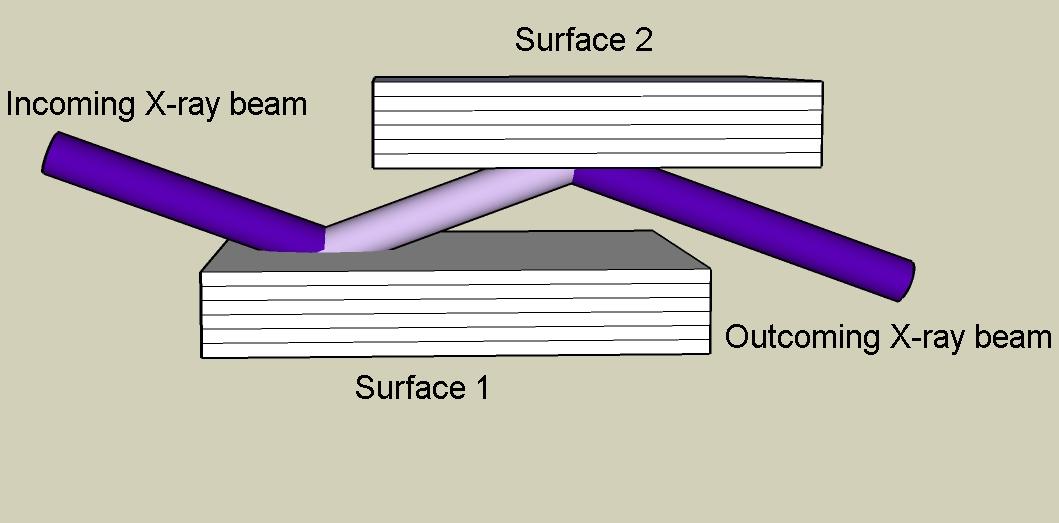}\\
\caption{\label{fig:bragg} Scheme of the proposed alignment procedure. An X-ray beam is impinging on the first surface at the Bragg angle condition. The X-ray beam will be diffracted again only if the second surface is parallel to the first one.}
\end{figure}

\begin{figure}[b]
\center
\includegraphics[width=0.95\columnwidth]{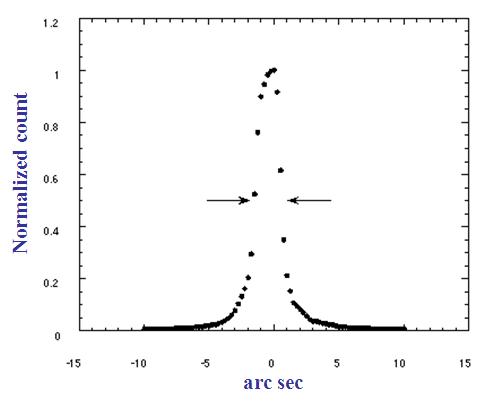}\\
\caption{\label{fig:rock} Rocking curve for Silicon (3 3 3)}
\end{figure}

The level of precision needed for measuring interaction forces in the plane-plane geometry requires also a particular attention on the insertion of a force detection system into a plane/plane set-up with movable surfaces. In the case of sphere-plane measurements, this problem is generally solved by gluing a sphere at the extremity of an AFM cantilever \cite{Mohi, Guillaume, Cap1, Cap2, NatPhot} . In the case of the plane-plane geometry, we propose to use a FIB in the realisation procedure. The FIB allows us to combine the need for a flat and lattice-oriented surface together with the insertion of a deformable lever mechanically linked with this oriented surface so that interaction forces can be measured. Thanks to a FIB equipped with an \textit{in} situ micromanipulator, a cubic like block can be cut and extracted from a wafer and welded at the end of a cantilever (see figure \ref{fig:fib}: in this particular case the block has been cut from a silicon wafer). During the positioning of the block, the precision in the angle that can be achieved is in the order of $10^{-1}$ $\deg$. Furthermore, the block surface can also be polished using FIB to obtain a better quality of the surface (roughness less than 10 nm r.m.s.).
\\
The positioning obtained after the insertion of the two surfaces and the force measurement system is not better than 0.1 $\deg$. This is not enough for interaction measurements in the plane-plane geometry. Beside changing and controlling the gap size, it is also necessary to change and control the relative orientation of the two surfaces. To this purpose, the two interacting surfaces are put on different mechanical stages that have to be approached each other. The approach can be performed using a linear translator based on inertial motors. To achieve the required precision in parallelism, a two angle tilter system has to be implemented in the experimental set-up. Micro-goniometers, that have been recently developed, allow one to adjust angle with a precision, at room temperature, better than $\ 10^{-4}$ $\deg$ (see for example \cite{attocube}).\\
To control the parallelism between the two surfaces we propose here to use X-ray diffraction onto single crystal. Let us consider the case sketched in figure \ref{fig:bragg}: a X-ray beam can impinge onto the surface of the sample at an angle satisfying the Bragg conditions for the diffraction. The X-ray beam diffracted by the surface presents the same characteristics as the incoming one, in terms of energy and intensity (for a single crystal thickness $t>50\ \mu m$ the diffracted intensity is $I_D>95\%\ I_0$, with $I_0$ the incoming beam intensity; see \cite{Zacha}). The diffracted beam impinges then onto the surface of the second sample. The beam is diffracted again only if the second crystal is orientated so that the Bragg condition is fulfilled.\\ 
Considering the scheme in figure \ref{fig:bragg} we note that the second surface satisfies the Bragg condition only when it is perfectly parallel to the first surface (see below). This technique allows then to control the parallelism between surfaces by recording the evolution of the out-coming X-ray beam intensity as a function of the relative orientation  ($\theta$, $\phi$) between the interacting surfaces. The out-coming beam intensity reaches its maximum value when the lattice vectors of the two surfaces are parallel. Using a single information it is then possible to control both angles $\theta$, $\phi$. The precision that can be achieved using such a control procedure is given by the Rocking curve of the selected materials. Let us consider for example the case of Silicon (3 3 3). From figure \ref{fig:rock} we note that a precision in angle better than $10^{-5}$ $\deg$ can be obtained. Such an alignment procedure between two silicon single crystals is of common use for double crystal monochromator in synchrotron facilities. 
\\

\begin{figure}
\center
\includegraphics[width=1\columnwidth]{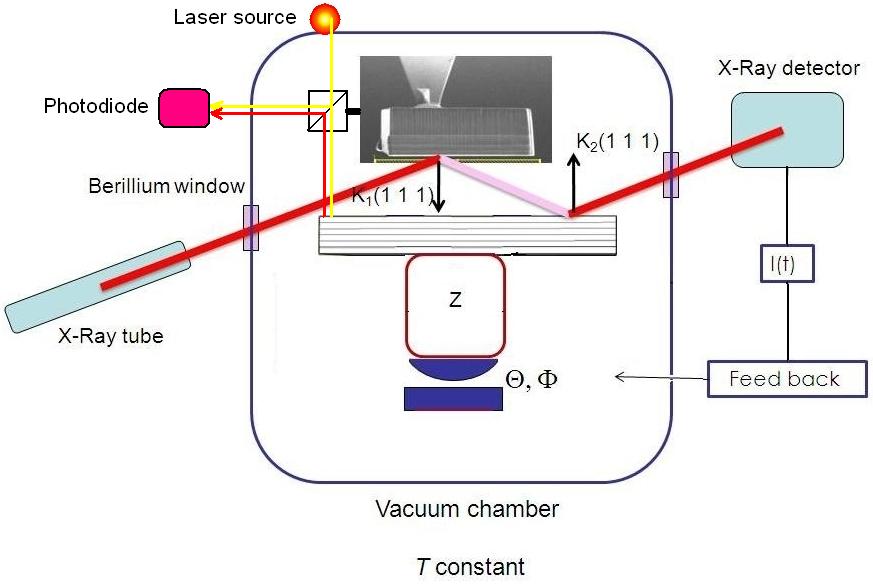}\\
\caption{\label{fig:set-up} Scheme of the proposed experimental set-up. A sample is mounted over a three-axis translation system and a two-angle tilt system. The probe, \textit{e.g.} a cubic like block attached to an AFM cantilever, is measured through a fibre-based interferometer.}
\end{figure}

Due to fluctuations in temperature, the two interacting surfaces continuously drift if there is no real-time control. The drift affecting the distance between the surfaces can be controlled using, first, thermalisation of the whole experimental set-up, and, on the top of that, real-time calibration using either electrostatic or optical measurement \cite{Charlaix1, Iannuzzi1, Iannuzzi2}. In this p2SFM an \textit{in-situ} real time optical interferometric measurement of the distance between the static lever basis and the movable single crystal must be implemented. The drift affecting the parallelism can be controlled as shown in figure \ref{fig:set-up}. A feedback loop acting on the goniometers can be related to the X-ray beam intensity detected. Maintaining constant the detected intensity ensures the real time parallelism of the two interacting surfaces.
The precision in the separation and parallelism between two surfaces that can be obtained by the here proposed p2SFM will allow one to make a reliable measurement of the interactions between two plane surfaces. The precision in the parallelism, in particular, is enough to measure also the interaction force that exhibits the strongest dependence on the distance, \textit{i.e.}, the Casimir force.\\
In summary, we have first recalled the performances in orientation control that are necessary to perform interaction force measurements in the plane-plane geometry at the nanoscale. In order to reliably design a plane-plane surface force machine, we have then proposed an original combination of existing elements originating from different fields of instrumentation. Beside classical thermalisation and distance control,  we shall use:
\begin{itemize}
\item combination of oriented flat surfaces and a lever for force measurements that is based on FIB, a key tool in nanotechnology (see fig. \ref{fig:fib});
\item nanopositioning that has been originally developed for scanning probe microscopy and that is based on inertial motors;
\item precise control of orientation based on X-ray diffraction on high quality single crystal that is routinely used at the required precision in synchrotron facilities.
\end{itemize}
Although up to now the set-up proposed here has not been realized and tested, in our view, its design will give birth to a new generation of versatile and original force machines dedicated to the investigation on non-contact interactions between surfaces in the plane-plane geometry and at the nanoscale.

\end{document}